\documentclass[preprint,12pt]{elsarticle}

\usepackage[utf8]{inputenc}
\usepackage{amssymb}
\usepackage{amsmath}
\usepackage{url}

\newcommand{\be}{\begin{eqnarray}}
\newcommand{\ee}{\end{eqnarray}}
\newcommand{\ba}{\begin{align*}}
\newcommand{\eea}{\end{align*}}
\newcommand{\hMpc}{{\ifmmode{h^{-1}{\rm Mpc}}
\else{$h^{-1}$Mpc}\fi}}

\journal{Astronomy and Computing}

\begin{document}

\begin{frontmatter}



\title{Ginnungagap -- a massively parallel 
cosmological initial conditions generator}


\author{Sergey Pilipenko} 

\affiliation{organization={AstroSpace Centre of P.N. Lebedev Physical Institute of the Russian Academy of Sciences},
            addressline={Profsojuznaja 84/32}, 
            city={Moscow},
            postcode={117997}, 
            state={},
            country={Russia}}

\author{Gustavo Yepes}
\affiliation{organization={Departamento de Física Teórica M-8 and CIAFF, Universidad  Autónoma de Madrid, Cantoblanco, 28049, Madrid, Spain}}

\author{Stefan Gottl\"{o}ber}
\affiliation{organization={Leibniz-Institut f\"{u}r  Astrophysik Potsdam (AIP), An der Sternwarte 16, D-14482 Potsdam, Germany}}

\author{Steffen Knollmann}
\affiliation{organization={Departamento de Física Teórica M-8, Universidad  Autónoma de Madrid, Cantoblanco, 28049, Madrid, Spain}}

\begin{abstract}
\texttt{Ginnungagap} is a fully parallel (MPI+OpenMP) code designed to generate cosmological initial conditions for simulations involving very large numbers of particles. It operates in several modes, including the creation of initial conditions with either uniform or spatially varying resolution (for “zoom-in” simulations). The initial conditions can be fully random or derived by extending the resolution of existing ones while preserving the large-scale structures. Ginnungagap is open source and modular, consisting of a collection of independent tools that can be used for a variety of tasks. In this paper, we describe the main features of Ginnungagap and present test results for different types of simulations prepared with it.

\end{abstract}




\end{frontmatter}



\section{Introduction}
Simulations of cosmic  structure formation from small density fluctuations 
are a fundamental tool in modern cosmology. The objects of interest span an enormous range of masses—from dwarf galaxies (< $10^7 M_\odot$) to superclusters (> $10^{16} M_\odot$)—and their study requires resolving both internal structures and large-scale environments. This corresponds to spatial scales ranging from sub-parsec to multi-gigaparsec regions. Modelling all these scales simultaneously within the volume of the observable Universe is not yet  feasible. Therefore, cosmologists always choose  a balance between the desired resolution and the available computational resources. Nowadays, typical  cosmological simulations involve between $10^8$ and $10^{13}$ particles and are executed on massively  parallel supercomputers with distributed memory architectures.

Modern codes used for running simulations, such as \texttt{Gadget-2} \cite{Springel05}, \texttt{Gadget-4} \cite{Springel21}, \texttt{Arepo} \cite{Arepo}, \texttt{RAMSES} {\cite{Ramses}, \texttt{GIZMO} \cite{GIZMO}, \texttt{ENZO} \cite{ENZO}, \texttt{SWIFT} \cite{SWIFT} and \texttt{PKDGRAV} \cite{PKDGRAV}}, are well optimized for distributed-memory supercomputers and show good scaling performance when using $10^4$ or more computational cores. However, until recently, there were no massively parallel and open-source initial conditions (IC) generators providing all of the following features simultaneously:

\begin{itemize}

\item {Zoom technique:} capability to create ICs with spatially varying resolution, with high-resolution (low-mass) particles in the region of interest and low-resolution (high-mass, fewer) particles filling the rest of the volume to properly reproduce large-scale perturbations.

\item {Resolution extension:} ability to increase the resolution of existing simulations. One can start with a low-resolution simulation, verify that the large-scale structures meet certain requirements, and then generate higher-resolution ICs preserving the same large-scale features while adding new small-scale perturbations.

\item {Efficient parallelization:} traditional IC codes using parallel Fast Fourier Transforms (FFTs) often divide the 3D domain into 1D slabs, limiting the number of cores to the grid size along one dimension. Dividing the volume into 2D subvolumes is far more efficient.

\item {Reuse of previous computations:} ability to reuse previously calculated data when producing multiple ICs on the same large-scale basis (e.g. when generating several zoom regions), thus avoiding redundant recalculation of the low-resolution components.

\end{itemize}

Some of these features are available in widely used public codes. \texttt{N-GenIC} \cite{Ngenic} is an MPI-parallel code, but it cannot generate zoom simulations nor refine the resolution of existing ones. The zoom technique is implemented in the \texttt{MUSIC} code \cite{Hahn2011}, which, however, is not MPI-parallel and is therefore limited in the total number of particles it can handle. {\texttt{MUSIC2-monofonIC} \cite{monofonic} is MPI+OpenMP parallel and precise (it includes many features, such as higher order perturbation theory, isocurvature perturbations, separate treatment of baryons, non-Cartesian particle lattices, etc.), but cannot make zoomed ICs.}

In this paper, we introduce \texttt{Ginnungagap}\footnote{The word Ginnungagap means ``primordial chaos'' in Norse mythology.}---an MPI-parallel code for generating large-scale, either zoom-in or homogeneous, cosmological initial conditions. It has been successfully used to create ICs for several major projects, including CLUES \cite{PIL_CLUES}, HESTIA \cite{PIL_HESTIA}, The Three Hundred Clusters \cite{PIL_300}, and SLOW \cite{PIL_SLOWI}.

\texttt{Ginnungagap} is publicly available on GitHub\footnote{\url{https://github.com/ginnungagapgroup/ginnungagap}} under the GPL-3.0 license. The repository includes comprehensive documentation with detailed installation and usage instructions. An automated installer script is also provided to download and compile all the required external libraries and dependencies needed to build and run the code.

The structure of this paper is as follows. In Section~\ref{sec:paradigm}, we describe the conceptual design of the code. Section~\ref{sec:tests} discusses test results for two typical use cases. 
Finally, Section~\ref{sec:summary} gives a brief summary of the paper. The mathematical formulation of the resolution refinement algorithm is presented in ~\ref{sec:wn}.

\section{ICs Preparation Paradigm}\label{sec:paradigm}
The preparation of initial conditions (ICs) in \texttt{Ginnungagap} is divided into three main steps:
\begin{enumerate}
\item Generation of the white noise (WN) field, which is a grid of independent Gaussian random numbers with zero mean and unit variance.
\item Computation of the velocity fields (VFs)—three grids corresponding to the three velocity components. If required, the density field is also generated.
\item Computation of particle positions and output generation in the appropriate format for running cosmological simulations.
\end{enumerate}

These steps differ slightly depending on the type of ICs being produced. For a single-resolution (non-zoom) simulation with random ICs, step~1 simply involves generating random numbers. For a resimulation of existing ICs at a different resolution, the WN and VFs are computed with constraints derived from the previous realization. In the case of zoom-in simulations, the WN and VFs must be generated at several resolutions, and a mask defining the zoom region must be applied in step~3. The position of the zoom region may also be required during steps~1 and~2 to reduce the memory footprint of the WN and VF grids.

The organization of \texttt{Ginnungagap} is shown in Table~\ref{tab:ggp}. The WN and VFs can be considered as temporary results not typically needed by the user of the ICs, but \texttt{Ginnungagap} stores them in files to reduce memory overhead. This greatly reduces memory requirements, since the code can work on one velocity component at a time. This also allows for the reuse of the data; for example, when creating multiple zoom regions from the same parent box. When the VFs exist, creating zoomed ICs requires only a small fraction of memory and computational time required for making VFs. E.g. in `The Three Hundred Clusters' project at the base resolution of $3840^3$ {the creation of the VFs took around 100 core-hours and the subsequent creation of 300 zoom regions from existing VFs took less than 10 core-hours ($\sim$2 core-m/cluster). Re-generating the VFs for each of the clusters would result in $\sim 30$k core-hours ($\sim100$ core-h/cluster) which would be a waste of time. This high gain is achieved because the code doesn't need to read the whole VFs to memory -- it only reads the part covering the zoom region, which is fast for small regions.} However, fields can consume a lot of disk space, so users must decide whether to retain them.

The source code of \texttt{Ginnungagap} contains several internal libraries which implement operations on 3D grids and particle collections: {parallel FFT, parallel I/O, power spectrum and statistics calculation, etc. Our realization of the parallel FFT uses 2D MPI subdomains which allows more effective scaling in comparison with the 1D slabs: up to $N_1^2$ MPI tasks can be used, where $N_1$ is the 1D grid size. This approach is similar to, e.g., \texttt{P3DFFT} library \cite{P3D}. The code supports both single and double precision.}

\begin{table}[]
    \centering
    \begin{tabular}{|p{5cm}|p{7cm}|}
    \hline
        Tool & Operation \\ \hline
        \texttt{ginnungagap} & Creating random WN fields and creating VFs from the WN \\ \hline
        \texttt{realSpaceConstraints} & Rescaling the WN \\ \hline
        \texttt{refineGrid} & cutting, adding and interpolating 3D grids (e.g. VFs) \\ \hline
        \texttt{generateICs}, \texttt{graficCoord} & Creating files in GADGET-2 or GRAFIC formats \\ \hline
        \texttt{prepare\_ini.sh} & Preparing workflows for complex ICs \\ \hline
        \texttt{LareWrite} & Creating masks for zoom-in ICs \\ \hline
        \texttt{fileTools} & Checking and analyzing ICs \\ \hline
    \end{tabular}
    \caption{Composition of \texttt{Ginnungagap}}
    \label{tab:ggp}
\end{table}

In what follows (Sections ~\ref{sec:vf}, \ref{sec:genics}, \ref{sec:incr})  we explain the ICs creation steps in detail.

\subsection{Computing the density and velocity fields}
\label{sec:vf}
The density and velocity fields in \texttt{Ginnungagap} are computed by convolving the white noise field with the transfer function using the Fourier transform:
\be
\widetilde{\delta}(\mathbf{k}) = \sqrt{P(|\mathbf{k}|)}\;\widetilde{W}(\mathbf{k}),
\label{eq:dens_from_wn}
\ee
where $\widetilde{\delta}(\mathbf{k})$ is the Fourier transform of the density field $\delta(\mathbf{x})$, $P(k)$ is the cosmological power spectrum, $\widetilde{W}(\mathbf{k})$ is the Fourier transform of the real space white noise field $W(\mathbf{x})$. This method is widely used for the creation of the initial conditions, e.g. in \citep{Bertschinger2001}.

A slightly modified approach is used in several codes \citep{Salmon1996,Pen1997,Sirko2005,Hahn2011} which takes into account the shape of the simulation box by convolving the power spectrum with the box window function. This results in an accurate reproduction of the real space correlation function up to the scale $\sim L_{box}/2$ and also ensures that the normalization of the matter power spectrum of fluctuations,  given by the  $\sigma_8$ parameter\footnote{$\sigma_8$ is the dispersion of the density fluctuation field when it is integrated  in spheres with radius 8~Mpc/h.}, within the box is exactly equal to the cosmological value  regardless of the box size, provided that  it is larger than $32$~Mpc/$h$.  The box size needs to be at least 4 times larger than the radius of the sphere, according to \cite{Pen1997}. However, this modification distorts the small scale power spectrum. Ignoring this correction makes $\sigma_8$ in the simulation box systematically lower than the cosmological  one (by $\lesssim 5$\% for a 100~Mpc/h box, as can be seen in  Fig.~1 of \cite{Pen1997}). This correction will be implemented as an additional option in future versions of \texttt{Ginnungagap}.

The velocity fields are computed from the Fourier transform of the density field:
\be
\widetilde{v_j}(\mathbf{k}) = i \widetilde{\delta}(\mathbf{k}) {k_j \over |\mathbf{k}|^2},\;\;
j = (1,2,3).
\label{eq:vel_from_wn}
\ee

Computations of equations (\ref{eq:dens_from_wn}), (\ref{eq:vel_from_wn}) are performed inside the \texttt{ginnungagap} tool, which uses a WN file as input or can generate a new random white noise using the high-quality random number generator \texttt{SPRNG} library\footnote{\url{http://www.sprng.org/}} version 2.0. The tool \texttt{ginnungagap} also needs to know the cosmological parameters and the power spectrum in tabulated form. By default, it produces the VFs files (one per component) and prints the basic statistics  of the WN and VFs fields. It can also optionally output the density field and the  power spectrum computed from it. {The memory required to generate three VFs is slightly more than twice the size of the single grid in memory.}

By default, the code uses the \texttt{HDF5} library\footnote{\url{https://www.hdfgroup.org/solutions/hdf5/}} for parallel input and output of the WN and VFs. Alternatively, it supports the \texttt{GRAFIC} format, which can be easily read using Fortran and does not require the installation of \texttt{HDF5}.

\subsection{Creating the  particle phase-space representation}
\label{sec:genics}
The next step in constructing the initial conditions consists of generating the phase-space coordinates of the particles. This is done by applying the Zel'dovich approximation \cite{Zeldovich70} to the generated VFs:
\begin{equation}
    x_j = q_j - \frac{D}{a\dot{D}}v_j,
\end{equation}
where $x_j$ is the comoving particle coordinate along axis $j$, $q_j$ is the Lagrangian coordinate, $D$ is the linear theory growth factor, $\dot{D}$ is its time derivative and $a$ is the scale factor, $v_j$ are velocity components at the centre of each VF grid cell. It is planned to add support for the 2nd order Lagrangian perturbation theory (2LPT) in future releases.  Currently one should use a starting redshift in the range $z=90-150$ to reduce artifacts from using the Zel'dovich approximation. 

Currently \texttt{Ginnungagap} supports the creation of all kinds of ICs in GADGET format (called ``SnapFormat 1'' in \texttt{GADGET-2}) using the tool called \texttt{generateICs}, and also non-zoomed ICs in GRAFIC format with the \texttt{graficCoord} tool. 

If gas particles are also needed, they can be generated from DM particles by assigning additional position offset of 1/2 cell size in each of three dimensions, and then both DM and gas particles are moved once more to maintain the center of mass of each DM+gas pair at the position of the original DM particle. This is needed to ensure consistent objects positions between DM only and DM+gas simulations.

\texttt{generateICs} can perform a periodic shift of the particle coordinates, which may be needed for zoomed simulations to place the zoom region in the box center. It also assigns IDs to particles and can do this in two ways: either sequential IDs, which are available in non-parallel runs of \texttt{generateICs} only, or coordinate-based IDs, which allow to calculate particle's Lagrangian position from its ID, even for the zoomed simulation.

{\texttt{generateICs} does not need to store the whole VFs in memory, and can operate on small patches at a time, but it is MPI+OpenMP parallel for high performance on large particle numbers. In principle, \texttt{generateICs} can be used to make zoomed ICs from the VFs made by other codes, e.g. \texttt{MUSIC2-monofonIC}. In order to use 2LPT or 3LPT, however, \texttt{generateICs} needs to use not only VFs, but also displacement fields produced by \texttt{MUSIC2-monofonIC}. This is a small modification of \texttt{generateICs} and we plan to implement it in future.}

Particle representations in formats other than \texttt{GADGET} can be obtained by converting the GADGET output files. However, initial conditions (ICs) in \texttt{GRAFIC} format---used by codes such as \texttt{RAMSES} and \texttt{ENZO}---can be generated directly by \texttt{Ginnungagap} for non-zoom simulations. In this format, the ICs consist of seven files: three for the velocity field components, three for the particle coordinates, and one for the density field.

To produce these files, the user must first configure \texttt{Ginnungagap} to output the velocity fields (VFs) in \texttt{GRAFIC} format instead of \texttt{HDF5}, and to enable the output of the density field. The VFs and density field produced by \texttt{Ginnungagap} can then be used directly as part of the ICs. In the second step, the auxiliary tool \texttt{graficCoord} generates the particle position files from the VFs.

\subsection{Increasing or reducing the resolution}
\label{sec:incr}

The workflow for increasing the resolution of an existing simulation consists of the following steps:
\begin{enumerate}
 \item Increasing the resolution of an existing low-resolution white noise (WN) field by applying real-space constraints.
 \item Computing the velocity fields (VFs) for both the low- and high-resolution realizations.
 \item Combining the small-scale Fourier modes from the high-resolution VFs with the large-scale modes from the low-resolution fields to obtain the final VFs.
 \item Generating the particle representation from the VFs using the Zel'dovich approximation.
\end{enumerate}

Steps (2) and (4) are performed exactly as for random initial conditions (ICs), as described in Sections~\ref{sec:vf} and~\ref{sec:genics}. The new procedures introduced in steps (1) and (3) are detailed below.

The resolution of the white noise field is increased by generating a new WN realization from the existing one while applying constraints that preserve the large-scale structure of the original field. The algorithm implementing this procedure is described in ~\ref{sec:wn} and is available in the tool \texttt{realSpaceConstraints}. With this tool, the one-dimensional resolution of the WN grid can be increased by a factor of 2 or 3/2, or by any multiple of these factors through repeated application of the procedure.

The constraints on the WN are applied locally. For example, if the resolution is doubled in one dimension, the algorithm replaces every two cells of the low-resolution WN with four new cells in the high-resolution WN. In the case of a 3/2 increase, two cells are replaced by three. As a result, all Fourier modes are affected, including those already present in the low-resolution WN{, see Fig.~\ref{fig:WN_comparison} in the Appendix}. Our tests have shown that this may distort the large-scale structure {(see details in Section~\ref{sec:algorithms})}, so it is often desirable to reduce the influence of the rescaled WN on the large-scale modes. This can be achieved by combining the VFs obtained from both the low- and high-resolution WN fields.

The small scale modes are cut from the high-res velocity field by applying a spherical top hat filter in k-space:
\be
\label{eq:v_small}
\widetilde{v_j}^{small}(\mathbf{k}) = \widetilde{v_j}^{high-res}(\mathbf{k}) (1-f(|\mathbf{k}|r_s)),\, f(x) = (x<1)? 1 : 0
\ee
where $r_s$ is the cutoff scale. The large scale modes are cut from the low-res velocity field:
\be
\label{eq:v_large}
\widetilde{v_j}^{large}(\mathbf{k}) = \widetilde{v_j}^{low-res}(\mathbf{k}) f(|\mathbf{k}|r_s).
\ee
The large and small scale modes are then added together, but since the large-scale modes are obtained on the low-res grid, they are first interpolated onto the high-res grid using cloud-in-cell (CiC) method (which also involves deconvolution of the field by dividing it by the Fourier transform of the interpolation kernel).

This methodology for constructing high-resolution VFs   further allows their generation not only across the full simulation box  but also within smaller subvolumes, which is particularly relevant for producing zoomed initial conditions. Since the high-resolution WN  field is constructed locally, the refined WN grid may be restricted to a limited region of the original volume. The corresponding VFs in this subvolume are then obtained by interpolating the low-resolution VFs and superimposing the small-scale perturbations derived from the new high-resolution WN field.

\begin{figure}
    \centering
    \includegraphics[width=0.5\linewidth]{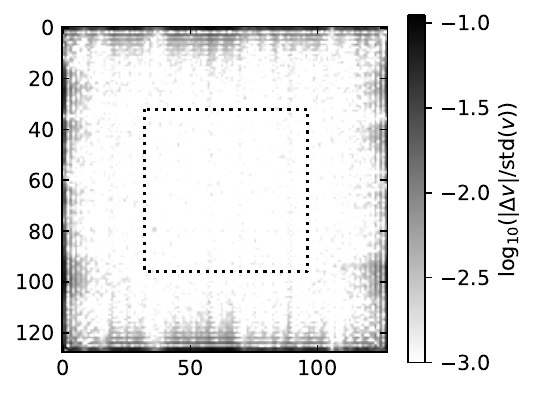}
    \caption{{Error of the velocity field inside a subvolume of 32~Mpc/$h$ size with $128^3$ grid cells inside a 64~Mpc/$h$ box. 
    The dotted frame indicates the recommended area for placing the zoom region.}}
    \label{fig:subbox}
\end{figure}

{The cutoff scale $r_s$ is chosen to be twice the low-res grid cell size. The CiC interpolation of the low-res VFs introduces an error with respect to the `perfect' Fourier-space interpolation (Whittaker-Shannon scheme). Increasing the cutoff scale $r_s$ reduces the impact of this error on the resulting VFs. On the other hand, increasing the cutoff scale increases the VFs error when the high-res grid covers a subvolume of the main simulation box.}
{This error arises due to the usage of the FFT to construct the VFs inside the subvolume, which applies the periodic boundaries inside the subvolume. The error is shown  in Fig.~\ref{fig:subbox}. The standard deviation of this error for our choice of $r_s$ is less than $10^{-3}$ at  distances larger  than $1/4$ of the subvolume size from its boundaries (inside the marked region in Fig.~\ref{fig:subbox}).}

The procedure for increasing the resolution is implemented through a sequence of calls to three tools:
\begin{enumerate}
    \item If working within a subvolume, \texttt{refineGrid} is first invoked to extract the subvolume from the low-resolution WN  and VF  grids (this step is skipped for the full box);
    \item \texttt{realSpaceConstraints} is then used to generate the high-resolution WN field;
    \item \texttt{ginnungagap} produces temporary small-scale VFs from the new WN by applying equations~(\ref{eq:dens_from_wn})–(\ref{eq:v_small});
    \item Finally, \texttt{refineGrid} combines the low- and high-resolution VFs by evaluating equation~(\ref{eq:v_large}) and adding the two fields.
\end{enumerate}

This workflow yields VFs covering either the full simulation volume or the selected subvolume.  
Full-box VFs can be used to produce single-resolution initial conditions (ICs) with \texttt{generateICs} or \texttt{generateGrafic}, whereas subvolume VFs are employed for generating zoomed ICs.

The resolution of existing ICs can also be degraded by applying nearest-grid-point (NGP) interpolation to a high-resolution VF and subsequently shifting it by half a low-resolution cell (via Fourier-space translation).  
This functionality is likewise implemented in the \texttt{refineGrid} tool.

\subsection{Zoomed initial conditions}
\label{sec:zoom}

Zoomed ICs in \texttt{Ginnungagap} consist of several particle sets with masses differing by factors of eight, corresponding to steps of two in one-dimensional grid resolution.  
These steps define the \emph{zoom levels}.  
Intermediate factors of $3/2$ are currently not supported between zoom levels.  
The highest resolution is applied to the region of interest, while the majority of the simulation volume is represented at the lowest resolution, with optional intermediate levels in between.

To construct such ICs, one requires the WN field from a full-box IC at some base resolution and a simulation output based on those ICs from which the objects of interest will be selected  for resimulation.  

The resolution of this base simulation typically lies between the highest and lowest zoom levels, although it may also coincide with the lowest resolution.  
An output snapshot of the base simulation is used to identify the Lagrangian region containing the target object (e.g., the halo to be re-simulated at higher resolution). 
This region, referred to as the \emph{mask} in \texttt{Ginnungagap}, is represented as a collection of marked cells of a grid at resolution of the base simulation. There is a tool which can be used to prepare a mask, called \texttt{larewrite}, but users can also provide their own masks. 
 In this tool, the input zoom region is defined as a collection of spheres within the base simulation at a given redshift, each specified by a center position and a radius. The centers may correspond to particular objects or spatial regions, and the radii should be sufficiently large to avoid contamination from massive low-resolution particles. Non-spherical regions can also be created by combining multiple overlapping spheres.

A grid matching the base IC resolution is then created, and particles located within the zoom spheres are traced back to their initial positions.  
Grid cells containing these particles are flagged, forming the mask. 
The \texttt{lareshow} tool can be used to visualise and inspect the resulting mask.

Masks produced in this way often contain \emph{holes} — unmarked cells surrounded by marked ones — which may lead to contamination by massive particles within the zoom region.  
To mitigate this, \texttt{larewrite} expands the initial mask by marking any cell that has more than a specified number of marked neighbours.  
The mask in \texttt{Ginnungagap} does not need to be a single connected structure.

In simulation codes such as \texttt{GADGET-2}, it is advisable to avoid having the zoom region intersect the simulation box boundaries.  
To achieve this, \texttt{larewrite} employs a clustering algorithm that determines optimal periodic shift parameters, placing the mask as far as possible from the boundaries.  
The user must then provide these shift parameters to \texttt{generateICs} when generating the final ICs.

Once the mask is defined, a set of VFs is created at the maximum, minimum, and (if required) intermediate resolutions by invoking \texttt{realSpaceConstraints}, \texttt{ginnungagap}, and \texttt{refineGrid}, as described in Section~\ref{sec:incr}.  
For each zoom level, the particle representation is then generated with \texttt{generateICs}, which also applies the mask to retain only the particles corresponding to that level.

When producing VFs for the various zoom levels (except the lowest), the user can choose between computing them on the full-box grid or within smaller grids (subvolumes) covering only the zoom region.  
Using subvolumes reduces both memory usage and disk space.  
With this approach, ICs with an effective local resolution equivalent up to $32,768^3$ particles in the full box have been successfully generated using \texttt{Ginnungagap}.

\subsection{Managing complex workflows}

While generating single-resolution ICs with \texttt{Ginnungagap} is relatively straightforward, producing zoomed ICs as described in Section~\ref{sec:zoom} may require multiple calls to \texttt{realSpaceConstraints}, \texttt{ginnungagap}, \texttt{refineGrid}, and \texttt{generateICs} in the correct order. Each of these tools is controlled by its own \texttt{.ini} file containing numerous parameters, and writing these files manually can be tedious. To automate this workflow, \texttt{Ginnungagap} provides a helper script called \texttt{prepare\_ini.sh}. This script takes a single “master” configuration file and generates all the configuration files required by the other tools. It also makes use of several machine-specific template files to produce ready-to-use job submission commands for various supercomputers on which the authors have tested the code. 

The \texttt{prepare\_ini.sh} script relies on the GNU \texttt{make} utility as a workflow manager, ensuring that all tools are executed in the correct sequence. Specifically, it produces a \texttt{Makefile} that encodes all dependencies between the \texttt{.ini} files, WN files, VFs, and the final Gadget-2 or GRAFIC outputs. After running \texttt{prepare\_ini.sh}, the user only needs to type \texttt{make gadget} or \texttt{make grafic} to run the full \texttt{Ginnungagap} workflow (at least on local machines). If any of the IC parameters are later modified, running \texttt{make gadget} again will only regenerate the intermediate files and IC levels affected by those changes.

\section{Code validation tests}\label{sec:tests}

Changing the resolution of a simulation—by adding or removing small-scale power—inevitably alters some of the properties of the simulated objects. When planning a series of simulations at different resolutions, it is useful to quantify the expected scatter in basic halo properties, such as mass and position, that arises from these resolution changes. In this section, we present results from several numerical experiments in which the same halos have been cross-identified across simulations with different resolutions.

Because of the complex shape of the cosmological matter power spectrum, the impact of resolution changes depends on the range of wavelengths included in the simulations and, consequently, on the halo masses being studied. We provide results for two representative halo masses: Milky Way–like halos ($\sim 10^{12}$~M$_\odot$) and galaxy cluster–like halos ($\sim 10^{15}$~M$_\odot$). {We also discuss how the impact of resolution depends on the algorithm and  code used, and compare \texttt{Ginnugagap} with the  well studied code \texttt{MUSIC2-monofonIC}.}

\subsection{Milky Way–mass halo in a 100~Mpc/$h$ box}
\label{sec:MWs}
A series of simulations of a Local Group (LG)–like pair of halos has been performed within the framework of the CLUES project. Details of the initial condition setup are described in \cite{PIL_CLUES}. The first simulation in the series was a full-box, low-resolution constrained simulation with $256^3$ dark matter particles in a box of size $L_{\mathrm{box}}=100$~Mpc/$h$. This simulation was used to identify the Lagrangian region corresponding to a 5~Mpc/$h$ radius sphere centered on the LG, which served as a mask for subsequent zoomed simulations. 

Zoomed initial conditions were then generated with effective resolutions corresponding to $N_{1D}^3 = 1024^3$, $2048^3$, $3072^3$, $4096^3$, and $8192^3$ particles in the full box. The ICs were generated at $z=99$ for the $4096^3$ and $8192^3$ resolutions, and at $z=80$ for the lower resolutions. In all zoomed setups, the outer regions of the box retained the base resolution of $256^3$ particles. The simulations were run using the \texttt{GADGET-2} code with the Tree+PM algorithm and employing the \texttt{PlaceHighResRegion} option to efficiently handle the zoom region.

\begin{figure}
    \centering
    \includegraphics[width=0.49\linewidth]{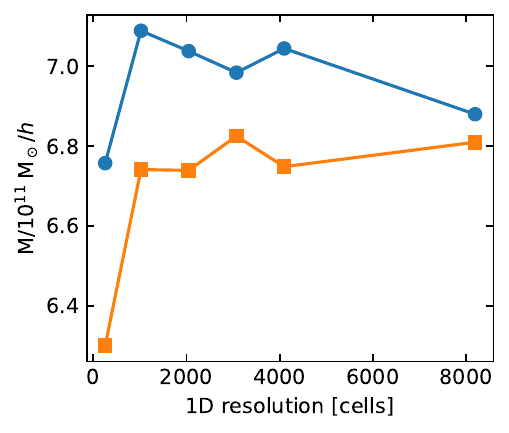}
    \includegraphics[width=0.49\linewidth]{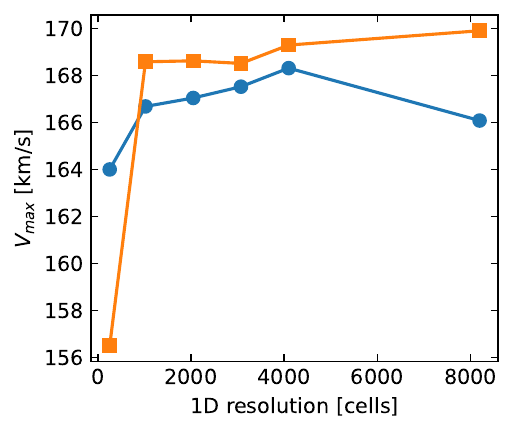}
    \includegraphics[width=0.49\linewidth]{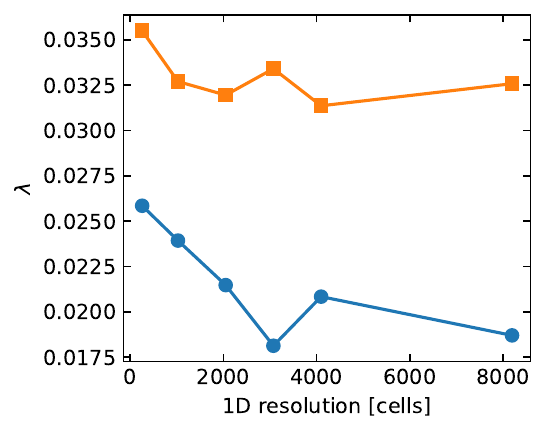}
    \includegraphics[width=0.49\linewidth]{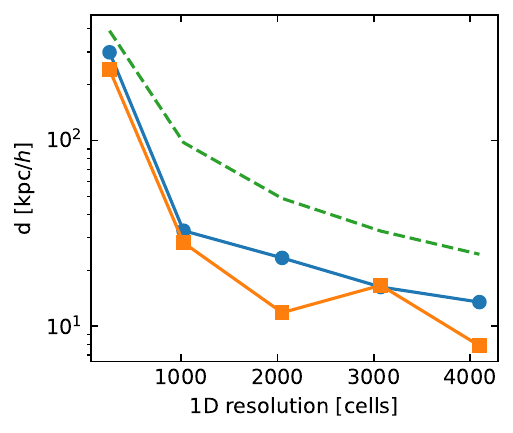}
    \caption{\textit{Top-left panel:} masses of the MW and M31 analogues (squares and circles, respectively) in the simulations with increasing resolution. {\textit{Top-right panel:} maximal circular velocities of MW and M31 analogues.} 
    {\textit{Bottom-right panel:} dimensionless halo spin parameters of MW and M31 analogues.} 
    \textit{Bottom-right panel:} distances of these two halos from their positions in the highest resolution ($8192^3$) simulation. Dashed line shows the ICs grid cell size $L_{box}/N_{1D}$. }
    \label{fig:LG}
\end{figure}

\begin{figure}
    \centering
    \includegraphics[width=0.49\linewidth]{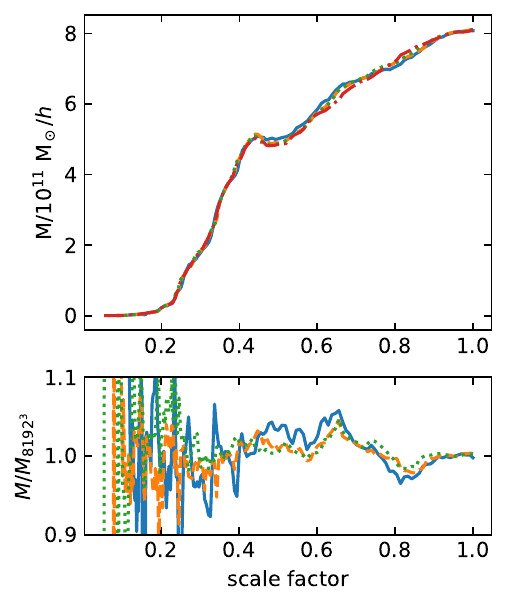}
    \includegraphics[width=0.49\linewidth]{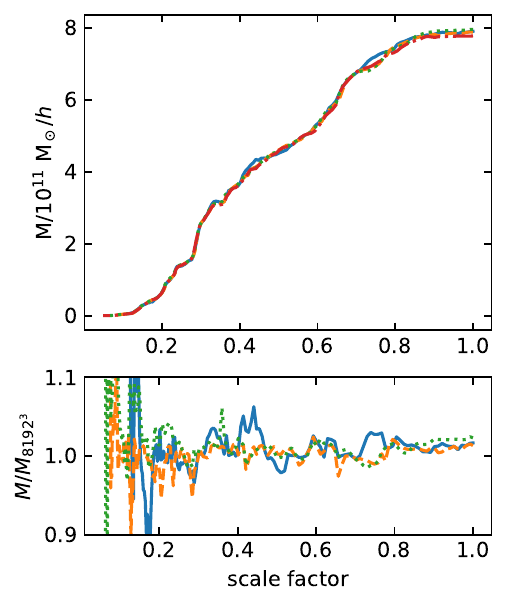}
    \caption{Top: mass accretion histories of the two MW-mass halos in zoomed simulations with effective resolutions $1024^3$ (solid lines), $2048^3$ (dashed lines), $4096^3$ (dotted lines) and $8192^3$ (dot-dashed lines) particles. Bottom: ratios of the main progenitor mass with respect to that in the highest resolution simulation.}
    \label{fig:mahs}
\end{figure}

We identify two halos resembling the Milky Way (MW) and M31 in these simulations at $z=0$ using the \texttt{AHF} halo finder \citep{AHF}, and we plot their masses (as determined by the halo finder) in the {top} left panel of Fig.~\ref{fig:LG}. The corresponding mass accretion histories (MAHs) of the two halos are shown in Fig.~\ref{fig:mahs}. For all zoom simulations ($N_{1D} \geq 1024$), the halo masses exhibit an RMS scatter of only about 1\% between different resolutions, with no significant systematic trend with resolution. In the non-zoomed $256^3$ simulation, each of the two halos contains roughly 100 particles, so the larger mass discrepancy relative to the higher-resolution runs is expected. The particle mass increases by a factor of 64 between the $256^3$ and $1024^3$ simulations. {The maximum circular velocity shows smaller variations than the virial mass (top right panel of Fig.~\ref{fig:LG}), while the halo spin parameter show larger variations (bottom left panel of Fig.~\ref{fig:LG}).}

The {bottom} right panel of Fig.~\ref{fig:LG} shows the displacement of the halo centers relative to their positions in the $8192^3$ simulation. The position errors are smaller than the size of a high-resolution cell (defined as $L_{\mathrm{box}}/N_{1D}$) for all simulations, with mean offsets of 0.5 and 0.4 cell sizes for the M31 and MW halos, respectively.

A similar approach for generating zoomed ICs has been employed in the HESTIA suite of simulations \citep{PIL_HESTIA}, which aim to reproduce Local Group analogues within a realistic cosmological environment. The main difference from the simulations described above is that the lowest-resolution part of the box was sampled with $64^3$ particles, and the runs were performed using the \texttt{AREPO} hydrodynamical code. The HESTIA suite includes three pairs of zoomed simulations of the same systems at two resolutions, $4096^3$ and $8192^3$. The masses of the Milky Way and M31 analogues in these pairs show an RMS scatter of about 2\% between the two resolutions.

\subsection{Cluster-mass halos in a 1~Gpc$/h$ box}
\label{sec:clusters}
\begin{figure}
    \centering
    \includegraphics[width=\linewidth]{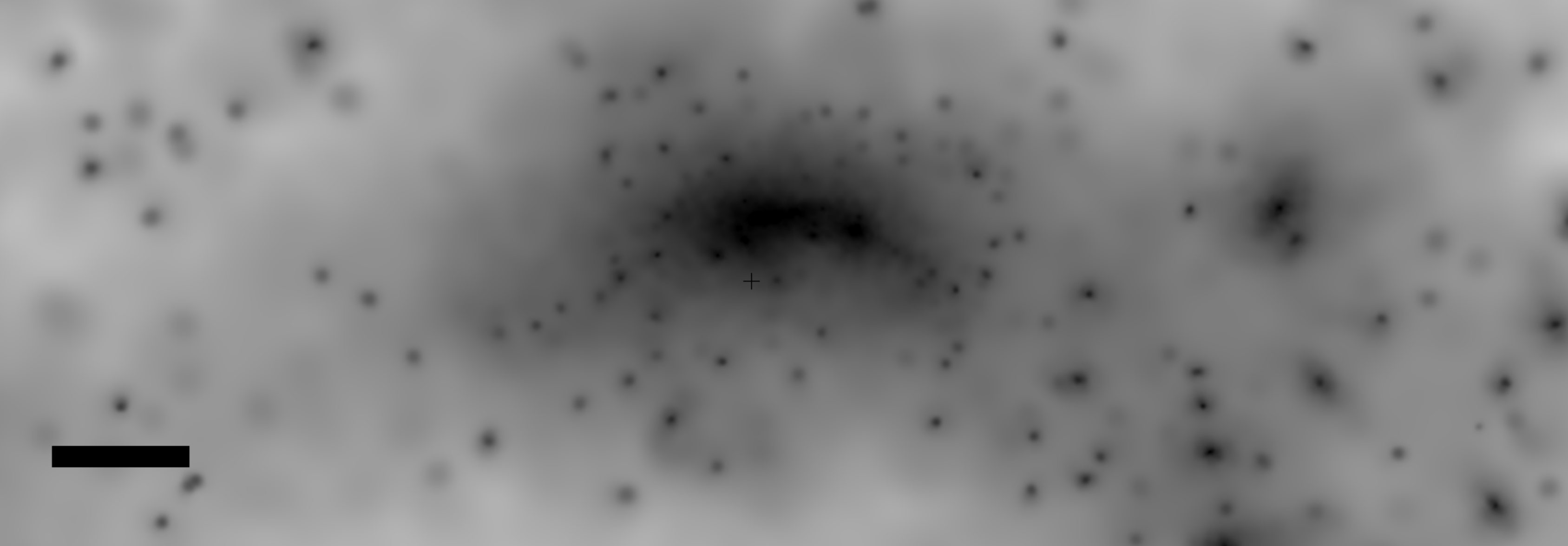}
    \includegraphics[width=\linewidth]{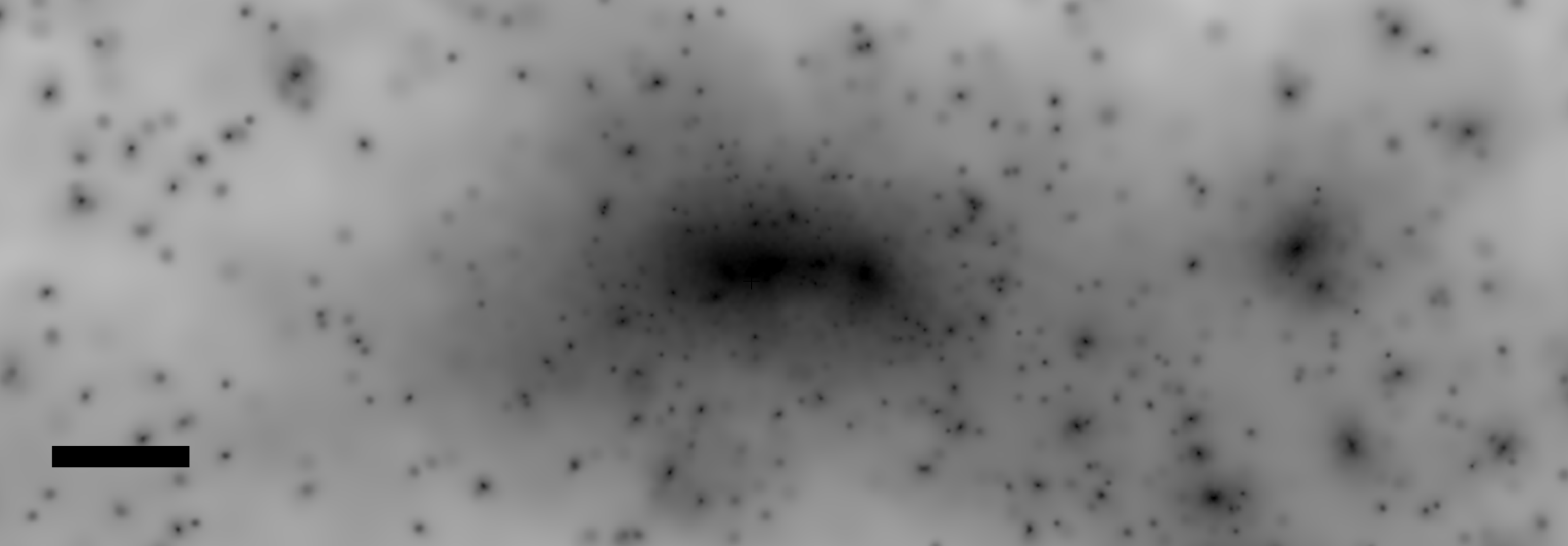}
    \includegraphics[width=\linewidth]{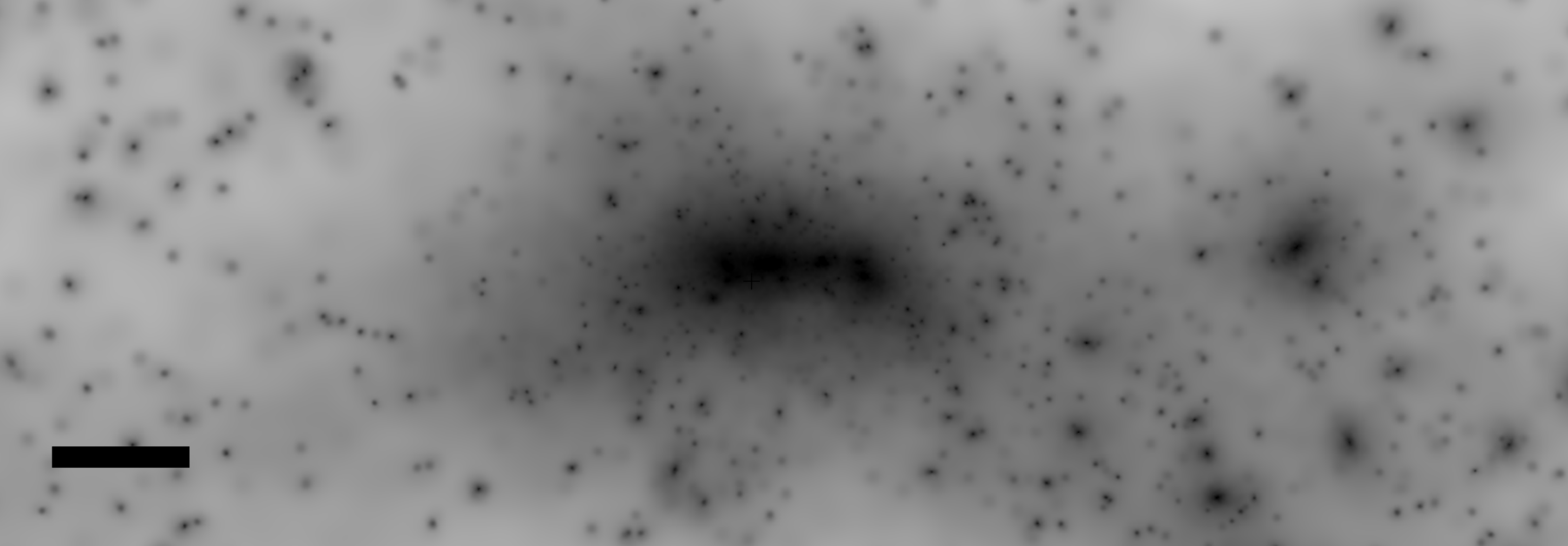}
    \caption{Projected DM density of a single cluster simulated with the zoom-in technique with the resolutions, equivalent to (from top to bottom) $3840^3$, $7680^3$ and $15360^3$ particles in the box. The black  bar on the left of the plots  corresponds to 1~Mpc/h scale}
    \label{fig:cluster}
\end{figure}

The \texttt{Ginnungagap} code was employed to generate the initial conditions (ICs) for \textit{The Three Hundred} project \citep{PIL_300}. This project aimed to re-simulate spherical regions centered on the 324  most massive cluster-mass halos, originally identified in the MultiDark Planck simulation (box size 1000~Mpc$/h$, $3840^3$ dark matter particles), using hydrodynamical codes at higher resolution. Since re-simulating the entire box would be computationally prohibitive, a separate zoomed IC was generated for each cluster. The first-generation ICs matched the resolution of the parent MultiDark simulation and were used for both dark-matter-only and full-physics hydrodynamical runs. Subsequently, higher-resolution zoomed simulations with an effective resolution of $7680^3$ particles were performed. Some of the clusters are being re-simulated with the resolution of $15360^3$ particles at the time of writing this paper. As an example, the DM density images for  one of the clusters at three different resolutions are shown in Fig.~\ref{fig:cluster}.

\begin{figure}
    \centering
    \includegraphics[width=0.49\linewidth]{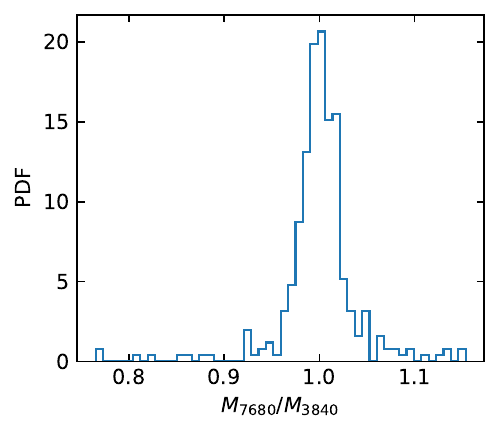}
    \includegraphics[width=0.49\linewidth]{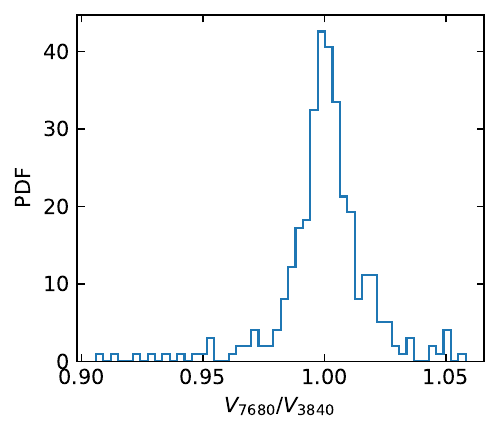}
    \includegraphics[width=0.49\linewidth]{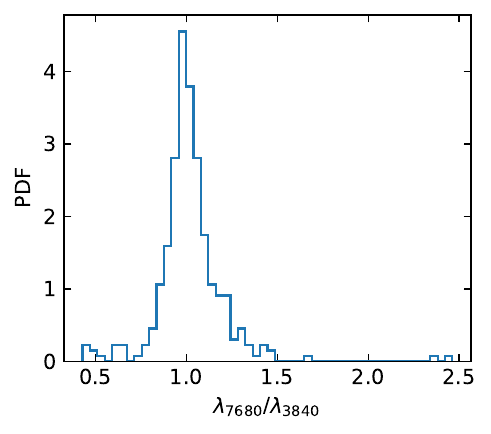}
    \includegraphics[width=0.49\linewidth]{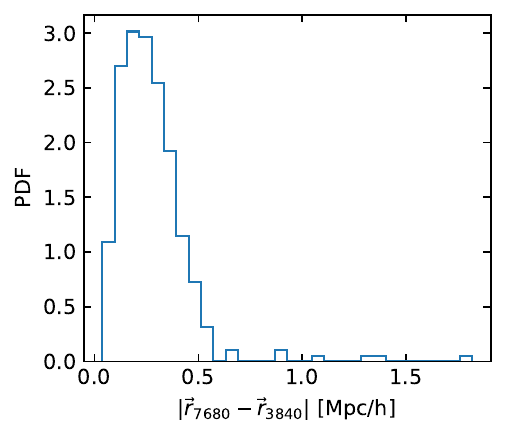}
    \caption{Probability distribution functions measured with 324 cluster-mass halos. \textit{Top-left panel:} mass ratio between $7680^3$ and $3840^3$ simulations. {\textit{Top-right panel:} maximum circular velocity ratio between $7680^3$ and $3840^3$ simulations.} 
    {\textit{Bottom-left panel:} dimensionless spin parameter ratio between $7680^3$ and $3840^3$ simulations.}
    \textit{Bottom-right panel:} distance between halo canters in $7680^3$ and $3840^3$ simulations.}
    \label{fig:pdf300}
\end{figure}

We also  compare the results of the dark-matter-only zoomed simulations for all the  324 clusters with effective resolutions of $3840^3$ and $7680^3$ particles within the 1000~Mpc$/h$ box. Halos in both simulation sets were identified using the \texttt{AHF} halo finder. The results are presented in Fig.~\ref{fig:pdf300}: the {top panels show the mass and the maximal circular velocity} differences between matched halos, while the {bottom panels display differences in the spin parameter and} offsets between halo centers in the two resolutions. The mean mass ratio and its $1\sigma$ scatter are $1.002 \pm 0.046$. {The mean maximum circular velocity ratio and  its $1\sigma$ scatter are $1.0002 \pm 0.019$. The mean spin parameter ratio and  its $1\sigma$ scatter are $1.012 \pm 0.21$.} The mean positional offset is 270~kpc$/h$, corresponding approximately to twice the high-resolution cell size. {The simulated halos contained from 2.5 to 14 million particles in the higher resolution runs.}

\subsection{How the resimulation algorithm affects halo properties}
\label{sec:algorithms}
{In 2016 we ran a series of tests which were used to develop the resolution extension algorithm for zoomed ICs, which is described in Section \ref{sec:incr} and \ref{sec:wn}. We used a single constrained realization of the WN in a 500~Mpc/h box and focused on the Virgo-like cluster with a mass of $4\times10^{14}$~M$_\odot$. We started from the $512^3$ particles simulation and measured the masses of this cluster in the zoomed resimulations with an effective resolution of $2048^3$. In the low resolution run the halo contained 5000 particles at $z=0$. We compared the virial mass of this cluster over last 10 snapshots covering a redshift range $z<0.35$ (last 4~Gyr).
}

{The $2048^3$ resimulations were made with five different algorithms:}
\begin{enumerate}
    \item {Hoffman-Ribak \cite{Hoffman1991} algorithm used to upscale the WN, as described by equation (\ref{eq:HR}). The halo virial mass changed by 11\% (averaged over last 10 snapshots) with respect to the $512^3$ simulation.}
    \item {WN upscaling described in \ref{sec:wn}. The mass changed by 3.8\%.}
    \item {WN upscaling described in \ref{sec:wn}, followed by copying the large-scale Fourier modes from the low-res fields. The mass changed by 1.8\%.}
    \item {WN upscaling followed by Fourier-space adjustments described by equations (\ref{eq:v_small})-(\ref{eq:v_large}) and Section~\ref{sec:incr}. The mass changed by 2.5\%.}
    \item {\texttt{MUSIC} code. The mass changed by 3.2\%.}
\end{enumerate}

{Thus, our simulations showed that the `local' updates of the WN without the restoration of the large scale modes gives the highest scatter, especially when the HR algorithm is used. The most `precise' method is copying the large scale Fourier modes from the low-res VFs while using small scale Fourier modes from the newly generated WN. However, this method requires the high resolution grid to cover the full box volume and hence the memory requirements become prohibitive for high resolution zoomed ICs. Our final algorithm produces scatter in halo masses comparable to that of \texttt{MUSIC} code.} 

{The authors of \cite{Onorbe14} studied the properties of halos in simulations based on initial conditions generated with the MUSIC code. They found that the maximum circular velocities of halos with $10^5$ up to few $10^6$ particles (in the highest resolution simulation) change by less than 5\% if the resolution has been  increased from $128^3$ to $512^3$ or from $256^3$ to
$512^3$.  Our results for the maximum circular velocity shown in Figs.~\ref{fig:LG} and \ref{fig:pdf300} are close to that finding. They also pointed out that the halo spin parameter is much more unstable with respect  to a similar increase of  resolution, namely
20\% changes for halos with $10^5$ up to few $10^6$ particles. Our results regarding the change of spins (Sections~\ref{sec:MWs} and \ref{sec:clusters}) are also on par with their results.}

{The work \cite{jenkins} reports sub-percent stability of halo parameters during resimulation with the \texttt{IC\_2LPT\_GEN} code \cite{Jenkins10} (not public available) and the \texttt{PANPHASIA} white noise fields. The Fourier grid resolution in these experiments varied from 3072 to 12288 cells in 1D, but the number of simulation particles was kept the same.}

\subsection{Comparison with \texttt{MUSIC2-monofonIC}}
{The code \texttt{MUSIC2-monofonIC} (for short, \texttt{monofonic}) is widely used in cosmological simulations and is well tested for the precision of the ICs it creates \cite{Hahn2011,Michaux21}. It is thus useful to check if the two codes produce identical ICs, given the same inputs. For that we generate the WN file with $128^3$ grid cells using \texttt{ginnungagap}, and then make the ICs in a 64~Mpc/h box with both codes with identical power spectrum. For that purpose we set up \texttt{monofonic} to use 1LPT (which is Zel'dovich approximation) and single precision. The resulting ICs are compared particle by particle, and we found that the standard deviation of velocities and displacements differences constitute $4\times10^{-7}$ of the standard deviations of those values themselves.}

{We also compare memory requirements and performance of the two codes. For that we generate ICs with $3840^3$ particles on the SuperMUC-NG supercomputer. \texttt{Ginnungagap} requires about 460 GB and fits into 6 thin nodes of SuperMUC-NG (48 cores and 96 GB physical memory per node). \texttt{monofonic} requires 660 GB and needed 9 nodes. As one can see,  the memory requirements are 40\% larger than  for   \texttt{Ginnungagap}.
The runtime of \texttt{Ginnungagap} was 1200 s on 6 nodes, which is equivalent to 96 core-hours. \texttt{monofonic} took 3930 s on 9 nodes or 470 core-hours. Thus, \texttt{Ginnungagap} turned out to be significantly faster with much less memory footprint, at least in the configuration used  in this test.}

\section{Summary and conclusions}
\label{sec:summary}

In this paper, we have {introduced} the cosmological initial-conditions generation code \texttt{Ginnungagap}. The main features of the code include efficient parallel performance, the ability to generate zoomed initial conditions, and the capability to modify the resolution of existing ones. {We summarise the basic features of various open source codes and compare them with \texttt{Ginnungagap} in Table~\ref{tab:codes}. \texttt{Ginnungagap} is } particularly well suited for producing high-resolution simulations of selected objects, for adding random small-scale perturbations to constrained realizations, and for supporting long-term projects in which simulation resolution is progressively increased in step with advances in supercomputing power.

We have also presented comparisons of the main properties of dark matter halos obtained from simulations with different resolutions, using initial conditions generated by \texttt{Ginnungagap}. The results demonstrate the robustness of the code in producing consistent halo masses and positions across a wide range of resolutions.

\begin{table}[t]
    \centering
    \begin{tabular}{|c|c|c|c|c|}
    \hline
     & \texttt{Ginnungagap} & \texttt{N-GenIC} & \texttt{MUSIC} &  \texttt{monofonic} \\ \hline
    Use existing WN & \checkmark & - & \checkmark & \checkmark \\ \hline
    Resolution extension & \checkmark & - & \checkmark & \checkmark\footnotemark[6] \\ \hline 
    Zoom & \checkmark & - & \checkmark & - \\ \hline
    MPI & \checkmark & \checkmark & - & \checkmark \\ \hline
    2LPT & -\footnotemark[7] & - & \checkmark & \checkmark \\ \hline
    3LPT\& precise baryons & - & - & - & \checkmark \\
    \hline
    \end{tabular}
    \caption{Main features of open-source cosmological initial condition generators.}
    \label{tab:codes}
\end{table}
\stepcounter{footnote}
\footnotetext{This can be done either for the \texttt{PANPHASIA} random realizations, or for the \texttt{MUSIC} non-MPI-parallel random number generator.} \stepcounter{footnote}
\footnotetext{Planned in future.}

\section{Acknowledgements}
The authors thank Noam Libeskind, Jenny Sorce, Klaus Dolag, Edoardo Carlesi and  Frazer Pearce, for useful discussions and the  extensive  use  and  testing of the code.
The authors gratefully acknowledge the Gauss Centre for Supercomputing e.V. (https://www.gauss-centre.eu/) for funding this project by providing computing time on the GCS Supercomputer SuperMUC-NG at the  Leibniz Supercomputing Centre (https://www.lrz.de/).

GY would like to thank  MICU/AEI/ (Spain) and FEDER/EU  for financial support under Project grants PID2021-122603NB-C21 and PID2024-156100NB-C21.

\appendix

\section{Increasing the Resolution of the White Noise Field}
\label{sec:wn}

Increasing the resolution of an existing simulation requires generating a new, higher-resolution white noise field that is not purely random, but constrained so that the large-scale structure remains identical in both realizations. Several methods can achieve this, such as padding in Fourier space, the Hoffman–Ribak (HR) algorithm \citep{Hoffman1991} (simplified for delta-correlated noise), their combination as implemented in \texttt{MUSIC} \citep{Hahn2011}, or the \texttt{PANPHASIA} method \citep{2013MNRAS.434.2094J}.

{\texttt{Ginnungagap} employs a modified version of the HR algorithm, which we describe here in one dimension for simplicity. 
The algorithm takes a small patch of size $N_L$ cells from the parent (low resolution) WN and replaces it with the new patch with size $N_H>N_L$. Thus, the resolution of the WN increases by the $N_H/N_L$ factor. The values inside a patch can be represented as an expansion over discrete basis functions $F_i(m)$, $m,i=0..N_L-1$ and $E_j(k)$, $k,j=0..N_H-1$. The basis functions must satisfy the orthonormality conditions:}
\begin{equation}
\sum_{k=0,N_H-1} E_i(k)E_j(k) = 
\begin{cases}
1, & i=j,\\
0, & i \neq j,
\end{cases}
\label{eq:ortonorm}
\end{equation}
{and the similar conditions apply to $F_i$.}

{In order to preserve the large scale structure, the first $N_L$ high-resolution basis functions $E_i$ must resemble $F_i$ basis functions. In that case, first $N_L$ expansion coefficients in the child patch just equal to the parent coefficients, while the rest $N_H-N_L$ coefficients are new and random. Since the $E_i$ and $F_i$ discrete functions are defined on different ranges, there is no strict one-to-one correspondence between them. We demonstrate how to construct the $E_i$ functions with an example of HR algorithm, in which, $N_L=1$ and $N_H=2$. The parent basis function is, evidently, $F_0(1)=1$. One can interpret $F_0$ as a constant function, consequently $E_0(m) = const$. Applying conditions (\ref{eq:ortonorm}) results in the following solution:}

\begin{flalign}
E_0 &= [1/\sqrt{2},\; 1/\sqrt{2}],\label{eq:E0}\\
E_1 &= [-1/\sqrt{2},\; 1/\sqrt{2}].\label{eq:E1}&&
\end{flalign}

{Thus, if $V$ is the value of the WN in a parent cell, and $R$ is the new Gaussian random number {with zero mean and unit variance}, the child cells are filled with values:}
\begin{equation}
    \left[ a_0E_0(0)+a_1E_1(0), a_0E_0(1)+a_1E_1(1) \right] = \left[ \frac{V-R}{\sqrt{2}}, \frac{V+R}{\sqrt{2}} \right],
    \label{eq:HR}
\end{equation}
{where $a_0=V$, $a_1=R$ are the expansion coefficients.}

{Our new algorithm has $N_L=2$ and $N_H=3$ or 4. We interpret $F_1(m)$ as a constant and $F_2(m)$ as a linear function of $m$, which results in the solutions described below.}

For the two parent cells:
\begin{flalign}
F_0 &= [1/\sqrt{2},\; 1/\sqrt{2}],\\
F_1 &= [-1/\sqrt{2},\; 1/\sqrt{2}].&&
\end{flalign}

For three child cells:
\begin{flalign}
E_0 &= [1/\sqrt{3},\; 1/\sqrt{3},\; 1/\sqrt{3}],\\
E_1 &= [-1/\sqrt{2},\; 0,\; 1/\sqrt{2}],\\
E_2 &= [-1/\sqrt{6}, \; 2/\sqrt{6}, \; -1/\sqrt{6}].&&
\end{flalign}

For four child cells:
\begin{flalign}
E_0 &= [1/2,\; 1/2,\; 1/2,\; 1/2],\\
E_1 &= [-\sqrt{7}/4,\; -1/4,\; 1/4,\; \sqrt{7}/4],\\
E_2 &= [-1/2,\; 1/2,\; 1/2,\; -1/2],\\
E_3 &= [1/4,\; -\sqrt{7}/4,\; \sqrt{7}/4,\; -1/4].&&
\end{flalign}

{If the two parent cells of the WN have values $V_0$, $V_1$, the first two expansion coefficients are $a_0=(V_0+V_1)/\sqrt{2}$, $a_1=(V_1-V_0)/\sqrt{2}$, and the rest coefficients $a_j$, $j>1$, are Gaussian random numbers with zero mean and unit variance. The values of the high-resolution WN patch are found as follows:}

\begin{equation}
    W(k) = \sum_{j=0,N_H-1} a_j E_j(k),
\end{equation}
{where equations (A.7)-(A.9) are used in case of $N_H=3$ and (A.10)-(A.13) are used in case of $N_H=4$. In particular, the high resolution WN patch for $N_H=3$:}

\begin{equation}
    \left[ \frac{V_0+V_1-R}{\sqrt{6}}-\frac{V_1-V_0}{2},
    \frac{V_0+V_1+2R}{\sqrt{6}},
    \frac{V_0+V_1-R}{\sqrt{6}}+\frac{V_1-V_0}{2}
    \right],
\end{equation}
{and for $N_H=4$:}
\begin{eqnarray}
\left[ \frac{V_0+V_1}{\sqrt{8}} -\sqrt{7}\frac{V_1-V_0}{4\sqrt{2}}-\frac{R_1}{2} + \frac{R_2}{4}, \right. \\
\frac{V_0+V_1}{\sqrt{8}} -\frac{V_1-V_0}{4\sqrt{2}}+\frac{R_1}{2} - \frac{\sqrt{7}R_2}{4}, \\
\frac{V_0+V_1}{\sqrt{8}} +\frac{V_1-V_0}{4\sqrt{2}}+\frac{R_1}{2} + \frac{\sqrt{7}R_2}{4},\\
\frac{V_0+V_1}{\sqrt{8}} +\sqrt{7}\frac{V_1-V_0}{4\sqrt{2}}-\frac{R_1}{2} - \frac{R_2}{4}
\left. \right] ,
\end{eqnarray}
{where $R$, $R_1$ and $R_2$ are Gaussian random numbers with zero mean and unit variance.}

{The choice of the coefficients in (A.10)-(A.13) is not unique. Since the functions $F_i(m)$ and $E_j(k)$ are defined on different domains, there is no straightforward correspondence between them. $E_0$ and $E_1$ must have the same symmetry around the middle of their domain as $F_0$ and $F_1$, respectively, otherwise the high-resolution WN will suffer from systematic shifts in one direction. An intuitive choice is that $E_0(k)$ is constant and $E_1(k)$ varies linearly with $k$ ($E_2$, $E_3$ are found from the orthonormality condition (\ref{eq:ortonorm})), i.e.:}

\begin{flalign}
E'_0 &= [1/2,\; 1/2,\; 1/2,\; 1/2],\\
E'_1 &= [-3/\sqrt{20},\; -1/\sqrt{20},\; 1/\sqrt{20},\; 3/\sqrt{20}],\\
E'_2 &= [-1/2,\; 1/2,\; 1/2,\; -1/2],\\
E'_3 &= [1/\sqrt{20},\; -3/\sqrt{20},\; 3/\sqrt{20},\; -1/\sqrt{20}].&&
\end{flalign}
{Another intuitive variant is that $E_1$ is the discretized sine function with the period equal to the patch size:}
\begin{flalign}
E''_0 &= [1/2,\; 1/2,\; 1/2,\; 1/2],\\
E''_1 &= [-1/2,\; -1/2,\; 1/2,\; 1/2],\\
E''_2 &= [-1/2,\; 1/2,\; 1/2,\; -1/2],\\
E''_3 &= [1/2,\; -1/2,\; 1/2,\; -1/2].&&
\end{flalign}

\begin{figure}
    \centering
    \includegraphics[width=0.7\linewidth]{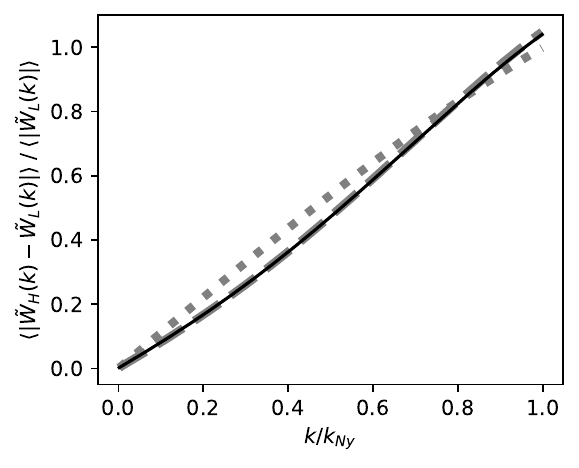}
    \caption{Error in the large scale part of the WN field when applying different rescaling algorithms: (A.10)-(A.13) (solid line), linear $E_1$ ((A.20)-(A.23)) (dashed line), HR algorithm and sine $E_1$ ((A.24)-(A.27)) (dotted line). $k_{Ny}$ is the Nyquist frequency of the low resolution WN.}
    \label{fig:WN_comparison}
\end{figure}

{We compared the difference of the high and low resolution WNs obtained using these variants for $10^6$ random realizations of 1D WN with 1024 cells in the lower resolution. The amplitude of the difference in Fourier space is shown in Fig.~\ref{fig:WN_comparison}. The lower is the value, the better is the reconstruction of the large scale modes in the high resolution WN. One can see that our initial variant (A.10)-(A.13) is almost identical to the `linear' variant (A.20)-(A.23). The first variant is actually 0.3\% lower on average than the `linear' variant, and 0.7\% lower at higher frequencies, $k/k_{Ny}>0.5$. Both the HR algorithm and the variant (A.24)-(A.27) give identical results, somewhat worse (by 16\% higher on average) than the two other variants. The difference in the WN between the first variant (A.10)-(A.13) and the `linear' one seems to be too small to have any impact on the simulated objects properties, but we keep the first variant for the compatibility reasons, since it was used in many simulations.}

{In three dimensions, the basis functions are constructed by multiplying the 1D basis functions, e.g.:}
\begin{equation}
E_{i,j,k}(m,n,l) = E_i(m) E_j(n) E_k(l).
\end{equation}
{To double the resolution, eight parent cells are first used to determine the eight low resolution coefficients, then 56 new coefficients are generated, and finally 64 values of the high-resolution white noise field patch are obtained.}

This approach is somewhat less accurate than \texttt{PANPHASIA}, since $F_1$ and $E_1$ do not exactly match, but it performs significantly better than the pure HR algorithm and is almost as simple to implement. An additional advantage is its flexibility: the resolution can be increased not only by a factor of two, but also by $3/2$ or $4/3$. Extending the set of $E_i$ functions beyond order 4 would, in principle, allow arbitrary fractional refinement factors.

The ability to refine a white noise realization in a consistent manner is crucial for multi-resolution or ``zoom-in'' cosmological simulations. It guarantees that the large-scale modes of the density field remain identical between the parent and refined simulations, preserving the cosmic environment of the region of interest. At the same time, the additional small-scale modes introduced through the high-resolution expansion provide new, independent fluctuations that seed finer structure formation. This ensures full statistical consistency across levels of refinement while maintaining computational efficiency and reproducibility.

\def\apj{Astrophys.~J}
\def\apjl{Astrophys.~J.,~Lett}
\def\apjs{Astrophys.~J.,~Supplement}
\def\an{Astron.~Nachr}
\def\aap{Astron.~Astrophys}
\def\mnras{Mon.~Not.~R.~Astron.~Soc}
\def\pasp{Publ.~Astron.~Soc.~Pac}
\def\aaps{Astron.~and Astrophys.,~Suppl.~Ser}
\def\apss{Astrophys.~Space.~Sci}
\def\ibvs{Inf.~Bull.~Variable~Stars}
\def\japa{J.~Astrophys.~Astron}
\def\na{New~Astron}
\def\aspproc{Proc.~ASP~conf.~ser.}
\def\aspcs{ASP~Conf.~Ser}
\def\aj{Astron.~J}
\def\actaa{Acta Astron}
\def\araa{Ann.~Rev.~Astron.~Astrophys}
\def\caosp{Contrib.~Astron.~Obs.~Skalnat{\'e}~Pleso}
\def\pasj{Publ.~Astron.~Soc.~Jpn}
\def\memsai{Mem.~Soc.~Astron.~Ital}
\def\astl{Astron.~Letters}
\def\aipproc{Proc.~AIP~conf.~ser.}
\def\physrep{Physics Reports}
\def\jcap{Journal of Cosmology and Astroparticle Physics}

\bibliographystyle{elsarticle-num}
\bibliography{cosmo}
\end{document}